\documentclass[sigconf]{acmart}
\usepackage{threeparttable} 
\AtBeginDocument{%
  \providecommand\BibTeX{{%
    \normalfont B\kern-0.5em{\scshape i\kern-0.25em b}\kern-0.8em\TeX}}}

\setcopyright{none}
\renewcommand\footnotetextcopyrightpermission[1]{} 

\begin{document}

\title[MultiPathGAN: Structure Preserving Stain Normalization using Multi Domain GANs]{MultiPathGAN: Structure Preserving Stain Normalization using Unsupervised Multi-domain Adversarial Network with Perception Loss}



\author{Haseeb Nazki}
\authornote{Corresponding Author}
\email{hn33@st-andrews.ac.uk}
\authornotemark[0]
\affiliation{%
  \institution{School of Computer Science, University of St Andrews}
  \city{St Andrews}
  \country{UK}
  \postcode{KY16 9AJ}
}

\author{Ognjen Arandjelović}
\email{oa7@st-andrews.ac.uk}
\affiliation{%
  \institution{School of Computer Science, University of St Andrews}
  \city{St Andrews}
  \country{UK}
  \postcode{KY16 9AJ}
}

\author{InHwa Um}
\email{ihu@st-andrews.ac.uk}
\affiliation{%
  \institution{School of Medicine, University of St Andrews}
  \city{St Andrews}
  \country{UK}
  \postcode{KY16 9AJ}
}
\author{David Harrison}
\email{david.harrison@st-andrews.ac.uk}
\affiliation{%
  \institution{School of Medicine, University of St Andrews}
  \city{St-Andrews}
  \country{UK}
  \postcode{KY16 9AJ}
}


\begin{abstract}
  Histopathology relies on the analysis of microscopic tissue images to diagnose disease. A crucial part of tissue preparation is staining whereby a dye is used to make the salient tissue components more distinguishable. However, differences in laboratory protocols and scanning devices result in significant confounding appearance variation in the corresponding images.
  This variation increases both human error and the inter-rater variability, as well as hinders the performance of automatic or semi-automatic methods. In the present paper we introduce an unsupervised adversarial network to translate (and hence normalize) whole slide images across multiple data acquisition domains. Our key contributions are: (i) an adversarial architecture which learns across multiple domains with a \emph{single} generator-discriminator network using an information flow branch which optimizes for perceptual loss, and (ii) the inclusion of an additional feature extraction network during training which guides the transformation network to keep all the structural features in the tissue image intact. We: (i) demonstrate the effectiveness of the proposed method firstly on H\&E slides of 120 cases of kidney cancer, as well as (ii) show the benefits of the approach on more general problems, such as flexible illumination based natural image enhancement and light source adaptation.
\end{abstract}




\maketitle

\pagestyle{plain}

\section{Introduction}
Digitization of pathological and histopathological slides in the medical imaging community has driven a significant increase in the development of computer aided diagnostic (CAD) systems~\cite{melo2020whole, daniel2011standardizing, webster2014whole}. One of the important steps in the process of tissue preparation for whole slide image (WSI) acquisition is staining, whereby histochemical stains such as hematoxylin and eosin (H\&E) are used to highlight the intensity of tissue slice components and thus make tissue structures distinct from each other~\cite{caie2021precision}. Staining facilitates the use of WSIs in manual or automated analysis by pathologists or CAD algorithms as a means of further the insight into the symptoms or mechanisms associated with a particular disease. However, batches of these slides collected from different groups or laboratories exhibit diverse stain styles due to high variability in staining protocol, dye manufacturers and scanners used to acquire these WSIs~\cite{bejnordi2014quantitative}. This variability may not greatly hinder the analysis of the tissue structure by an experienced pathologist; however, it significantly reduces the accuracy of existing CAD systems owing to their limited ability for generalization from data used to train them~\cite{tellez2019quantifying}. Meanwhile, given the massive amount of gigapixel-sized WSI data~\cite{dimitriou2021magnifying}, there is a growing demand to build fast, automated, and scalable pipelines for large-scale image analysis. In particular, normalization methods increase the accuracy of machine learning (ML) algorithms that use stain-normalized images as input to a pre-trained deep network~\cite{anghel2019high, mahapatra2020structure, gupta2019gan}.\par

Considering the importance of the problem, it is not surprising that a number of attempts at solving it have been described in the literature.
Simple attempts which focused on color matching~\cite{reinhard2001color,yue2019colorectal}, that is the color-channel alignment of a novel WSI with those of a reference template, can lead to unrealistic looking results as a consequence of the lack of learning over a representative corpus and the reliance on a single target image.
On the other hand, while stain-separation methods~\cite{daniel2011standardizing, macenko2009method} do consider each staining channel independently for normalization, they fail to take into  account the spatial features and the structure of the components in the tissue. Moreover, they rely on a skillfully chosen reference template image which can significantly affect the outcome~\cite{shaban2019staingan}. A related machine learning based approach employs a sparse autoencoder~\cite{janowczyk2017stain}, dividing an input image into multiple tissue regions and then independently normalizing each region using a single template. It can be readily seen that this approach suffers from the same limitation as that noted before, namely that a single reference template fails to capture the significant variation in the appearance of WSIs within a single lab.

More recently, deep learning-based approaches for stain normalization~\cite{shaban2019staingan, bentaieb2017adversarial, ke2021contrastive, runz2021normalization, kang2021stainnet, de2021residual, vasiljevic2021towards} using style transfer and image to image translation with adversarial learning methods~\cite{goodfellow2014generative, isola2017image, zhu2017unpaired, choi2020stargan} have made strides in first establishing and then repeatedly improving the state-of-the-art.
However, all of the existing adversarial network based methods are capable of translating between only two specific domains which makes them unsuitable for the inherently multi-domain challenge posed by the real-world clinical practice. Some of these approaches even rely on supervision, learning a mapping between the two domains given a labelled set of input-output pairs which are not always readily available, particularly for WSI datasets. Moreover, while these methods transfer the style content between the domains efficiently, they too struggle at preserving the structure of the tissue components. Previous efforts to develop algorithms capable of preserving the anatomical structure of tissue in a WSI while also translating the style between domains in an unsupervised setting have shown decent results~\cite{de2021residual, mahapatra2020structure, gadermayr2018way, lahiani2019perceptual}; nevertheless these methods too remain constrained by being able to deal with two specific domains only. Wagner et al.~\cite{wagner2021structure} use a GAN architecture based on disentangled representations~\cite{lee2020drit++}, which produces realistic looking high level structures but which poorly semantically correspond to the input.

In this work, we introduce MultiPathGAN, a unified deep learning adversarial network for translating histopathological WSIs obtained from different pathology labs/scanners across multiple domains for stain normalization while preserving the structure of the internal tissue components. 
The data adaptive nature of the MultiPathGAN means that it can also be readily employed in other applications wherein it is critical to constrain the effects of adversarial learning in a manner that retains the salient input image structure, while only translating the style content across multiple domains. This alleviates the problem of variation in the feature representation of training data by having a unified representation that can map many-to-many stain style domains instead of a simple one-to-one representation. Importantly, our network takes in training data from multiple domains, and learns the mappings between \emph{all of them} using only a single generator as proposed by Choi et al.~\cite{choi2018stargan}. However, unlike StarGAN~\cite{choi2018stargan} which translates images between multiple domains without effectively preserving the content of the WSI while changing the style-related part of the inputs, we also achieve the preservation of the fine salient anatomical structure using an auxiliary feature extraction network and perception loss to minimize the perceptual distance~\cite{ledig2017photo} between a real and the corresponding generated (synthetic), fake image. To summarize, we make the following key contributions:
\begin{itemize}
    \item We propose MultiPathGAN: an unsupervised multi-domain image translation network based on adversarial learning to normalize the stain variations between multiple WSI domains. 
    Training a single generator on multiple domains increases scalability and robustness as it allows the translation process to be steered towards any desired domain given its target label from any large set of available domain classes.

    
    \item To preserve the semantic information and capture the geometric and structural patterns of the source image, we introduce an additional feature extraction network in order to regulate multi-domain adversarial learning in a multi-domain setting. Consisting of a pre-trained convolutional classifier, the feature extractor aids the generator to optimize on the perceptual loss by providing perceptual distance between the extracted feature maps of the source and target WSI domains.
\end{itemize}

\section{Method}
In this section we first describe in broad strokes the proposed framework (MultiPathGAN) for multi-domain image-to-image translation aimed at preserving the structure of the salient input image content while effectively integrating it with the style content of the target. Following this coarse overview of the method, each of its constituent components is elaborated on in detail.

\begin{figure}[h]
\centering
\includegraphics[width=\linewidth]{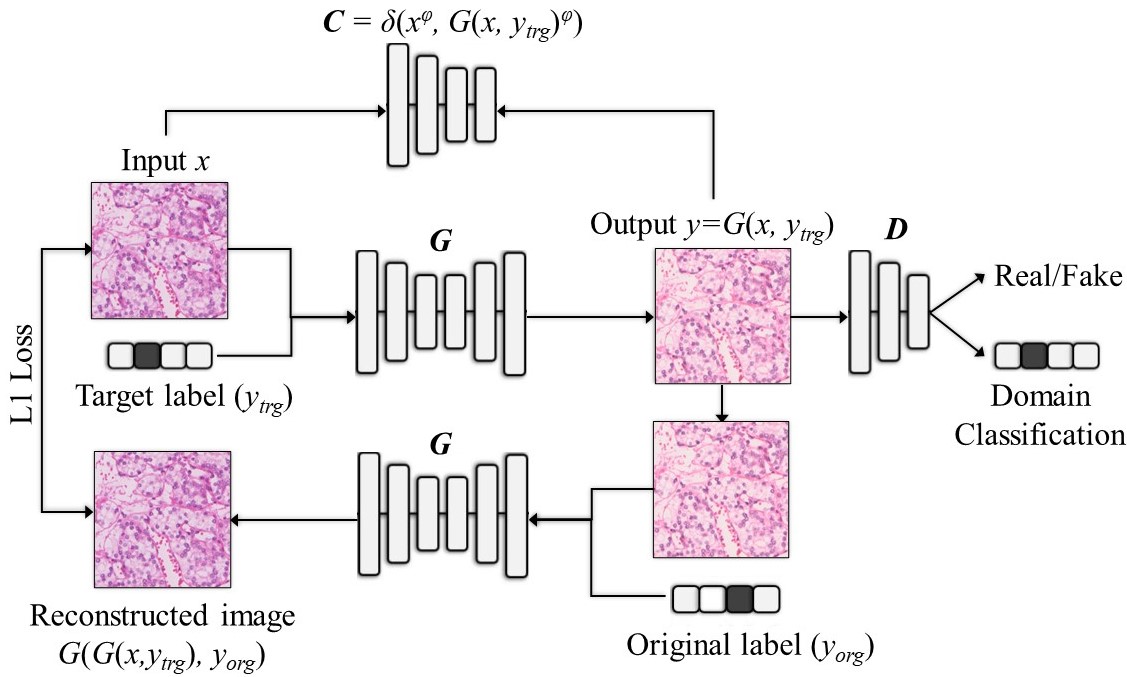}
\caption{Information flow in MultiPathGAN. Generator $G$ generates output image $y$, given input image $x$ and the randomly generated target domain label $(y_{trg})$. Discriminator $D$ distinguishes real from fake images, and simultaneously assures that $y$ is classifiable as being in the target domain. The input $x$ is also reconstructed from $y$ to enforce self-consistency. Concurrently, $x$ and $y$ are also passed through a common feature extraction network $C$ which provides feature maps for perceptual distance minimization.
}
\Description{figure description}
\label{fig:1}
\end{figure}

\subsection{MultiPathGAN}
One of the key ideas in MultiPathGAN lies in the use of an auxiliary feature extraction network. This network is used to integrate perceptual loss in a multi-domain adversarial network, thereby introducing a semantic relationship between the input and the output, and drives the network to learn how to retain the salient structural elements of the input. 

As illustrated in Fig.~\ref{fig:1}, MultiPathGAN trains a single generator $G$ to learn mappings between multiple domains using target labels. To achieve this, we train $G$ to translate an input image $x$ into an output image $y$ conditioned on the randomly generated target domain label $(y_{trg})$, such that $G(x, y_{trg}) \rightarrow y$. The target domain label $y_{trg}$ is generated randomly so that $G$ learns to translate the input image flexibly~\cite{choi2018stargan}. Further, an auxiliary classifier allows a single discriminator to control multiple domains by producing probability distributions over both sources and domain labels~\cite{choi2018stargan}. Note that the single generator is used twice, first to translate the original image into the target domain and then to reconstruct the original image from the translated image. Simultaneously, the input image $x$ and output image $y$ are fed to a pre-trained deep convolutional classifier $C$ to obtain feature maps $x^\phi$, $G(x, y_{trg})^\phi$ and obtain the mean squared distance between these activations~\cite{ledig2017photo}. This distance is used in an additional perception loss function to help preserve the perceptually subtle salient content in the input image as it is translated to the output domain~\cite{ledig2017photo, nazki2020unsupervised}. In the subsections which follow we provide detailed descriptions of all loss functions used in training the network that make this preservation possible.

\subsubsection{Adversarial Loss Function.}

As in previous work, $G$ tries to generate an image $G(x, y_{trg})$ conditioned on both the input image $x$ and the target domain label $y_{trg}$. At the same time, a discriminator $D$ tries to distinguish between real and fake images, $x$ and $G(x, y_{trg})$ respectively. To stabilize the training process and generate higher quality output of each mapping that matches the empirical distribution of the target domain, while focusing on the source domain, we adopt the Wasserstein adversarial loss~\cite{arjovsky2017wasserstein,magister2021generative} with gradient penalty~\cite{gulrajani2017improved}:
\begin{align}
  \label{eq1}
  \mathcal{L}_{Adv} & = \mathbb{E}_x[D_S(x)]-\mathbb{E}_{x,y_{trg}}[D_S(G(x,y_{trg}))] \notag \\ & \quad \quad \quad \quad \quad -\lambda_{gp}\mathbb{E}_{\hat{x}}[(\|\nabla_{\hat{x}}D_{S}(\hat{x}\|_{2}-1)^2] 
\end{align}
where $\lambda_{gp}$ is the coefficient for gradient penalty and $\hat{x}$ is uniformly sampled along a straight line between a real and a generated image pair~\cite{gulrajani2017improved}. $D_S$ refers to the probability distribution over the sources given by discriminator $D$.

\subsubsection{Domain Classification Loss Function.}

We also adopt the domain classification loss ~\cite{choi2018stargan} using an auxiliary classifier on top of $D$ when optimizing $G$ and $D$ to ensure that the input $x$ translated into an output image $y$ is properly classified to the target domain $y_{trg}$. This objective can be broken down into two terms: a domain classification loss calculated using real images to optimize $D$, and a domain classification loss calculated using fake images to optimize $G$:
\begin{align}
\label{eq2}
    \mathcal{L}_C^r=\mathbb{E}_{x,y_{org}}[-logD_{C}(y_{org}|x)]
\end{align}
\begin{align}
\label{eq3}
    \mathcal{L}_C^f=\mathbb{E}_{x,y_{trg}}[-logD_{C}(y_{trg}|G(x,y_{trg})]
\end{align}
where $D_C$ is the probability distribution calculated by $D$ over domain labels. In \eqref{eq2}, $D$ tries to minimize the objective by learning to classify a real image $x$ to its corresponding original domain $y_{org}$. At the same time, in \eqref{eq3}, $G$ tries to minimize the objective to generate images that could be classified as the target domain $y_{trg}$.

\subsubsection{Reconstruction Loss Function.}

To ensure that the translated images preserve the high level features of the input images while changing their domain specific characteristics, we apply a cycle consistency loss~\cite{zhu2017unpaired,choi2018stargan} to the generator $G$:
\begin{align}
\label{eq4}
    \mathcal{L}_{Cyc}=\mathbb{E}_{x,y_{trg},y_{org}}\|x-G(G(x,y_{trg}),y_{org})\|_1
\end{align}
where $G$ takes in the translated image $G(x, y_{trg})\rightarrow y$ and the original domain label $y_{org}$ as input and tries to reconstruct the original image $x$ using the $L1$ norm in the loss function.

\subsubsection{Perceptual Loss Function.}

In addition to the reconstruction loss used to retain the high level features, we also include a perceptual loss which preserves the fine structure of the input image during domain transfer as the perceptual distance is minimized~\cite{ledig2017photo,johnson2016perceptual} between the generated $G(x, y_{trg})$ and the original image $x$. This distance is calculated by feeding both the original and the transformed image to a pre-trained classification network $C$ and extracting their feature maps $x^\phi, G(x, y_{trg})^\phi$  respectively, at spatial resolution $\phi$:
\begin{align}
\label{eq5}
    \mathcal{L}_{P}=\delta(x^\phi,G(x,y_{trg})^\phi)
\end{align}
The mean squared distance is calculated for each feature map and the total perceptual distance is then a weighted sum of the $L2$ distances. By minimizing this distance, we implicitly compel the transformed image to contain the same structural meaning as the original image as perceived by a pre-trained classification network.

\subsubsection{Final Objective.}

The final objective used to train the discriminator $D$ and the generator $G$ can be written as follows:
\begin{align}
\label{eq6}
    \mathcal{L}_{D}=-\mathcal{L}_{Adv}+\lambda_{C}\mathcal{L}_{C}^{r}
\end{align}
\begin{align}
\label{eq7}
    \mathcal{L}_{G}=\mathcal{L}_{Adv}+\lambda_{C}\mathcal{L}_{C}^{f}+\lambda_{Cyc}\mathcal{L}_{Cyc}+\lambda_{P}\mathcal{L}_{P}
\end{align}
where $\lambda_{C}$, $\lambda_{Cyc}$ and $\lambda_{P}$ are the hyper-parameters to regulate the individual loss functions $\mathcal{L}_{C}$, $\mathcal{L}_{Cyc}$ and $\mathcal{L}_{P}$ respectively.

\subsection{Implementation}

\subsubsection{Network Architecture and Training.}


In our network, the generator $G$ is based on the six block ResNet~\cite{he2016deep} backbone and a pair of convolutional and transposed convolutional layers with the stride size of two on either side, respectively. We use instance normalization for the generator, which has been shown to improve performance~\cite{ulyanov2016instance}. For the discriminator $D$, we use PatchGANs~\cite{isola2017image} with no normalization and Leaky ReLU with a negative slope of 0.01 to classify the overlapping real and fake patches. Finally, for our feature extraction network $C$, we utilize a pre-trained 34-layer residual network~\cite{he2016deep}. We use the activations from the last convolutional layer of the ResNet feature extraction network to concentrate on the fine changes in these higher levels, propagated by its skip connections, which thus allows us to determine a better perceptual distance between the generated $G(x, y_{trg})$ and original image $x$.\par

For our experiments, we set the values of $\lambda_{gp} = 10$, $\lambda_{C} = 1$, $\lambda_{Cyc}=10$ and $\lambda_{P}=0.75$. We use a mini-batch size of 16 and a base learning rate of 0.0001 for both $G$ and $D$ which decays after every 10 epochs. Further, we use the Adam optimizer~\cite{da2014method} with $\beta_{1}=0.5$ and $\beta_{2}=0.999$ and following Gulrajani et al.~\cite{gulrajani2017improved}, we perform one $G$ update for every five $D$ updates over 80 epochs which takes about 21 hours to train our network on a single Nvidia DGX-1 instance.

\begin{figure}[h]
\centering
\includegraphics[height=15cm,width=7.2cm]{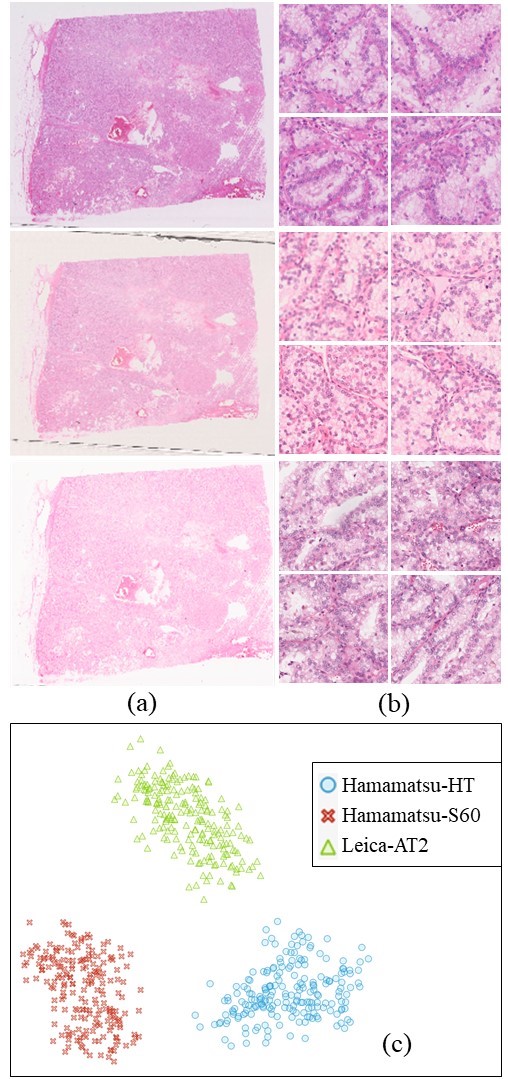}
\caption{(a) WSI samples from the three domains, (b) the corresponding patches, and (c) TSNE visualization of our dataset samples taken from the three WSI domains.}
\Description{figure description}
\label{fig:2}
\end{figure}

\section{Evaluation}

\subsection{Dataset.}

H\&E slides were prepared from a cohort of 120 cases of kidney cancer obtained from the Pathology Archives, Lothian NHS. Sections were cut at 3 $\mu$m and routinely stained for H\&E before being cover-slipped. Slides containing both cancerous and non-cancerous ``normal'' tissue were distributed to different laboratories and scanned locally on their scanners of choice before all images were collated at <removed for review>. Ethics approval was provided by Lothian NHS Biorepository (ES/15/0094). More specifically, we used WSI acquired from Hamamatsu NanoZoomer S60 at $20\times (0.42{\mu}m/pix)$ magnification, Hamamatsu NanoZoomer 2.0-HT at $20\times (0.45{\mu}m/pix)$ and Leica Aperio AT2 at $20\times (0.50{\mu}m/pix)$ magnifications, respectively, to create three dataset domains. The WSIs were then split into $256 \times 256$ non-overlapping patches. To train our MultiPathGAN, we used 10 WSIs for each domain and randomly extracted 1250 training and 200 testing sample patches for each class. We also used an additional domain of WSIs from Hitachi HV-F202SCL at $20\times (0.22{\mu}m/pix)$ magnification, unseen to the network at the training time. This domain was used to evaluate the generalization of our network and its ability to adapt to the high variance data space of WSIs in histopathological studies.

Fig.~\ref{fig:2} shows examples of WSIs and their respective tiles from our dataset. Even though the variation in the acquired data is readily apparent to the naked eye, we corroborate this in Fig.~\ref{fig:2}(c) using t-Distributed Stochastic Neighbor Embedding (t-SNE) visualizations, which help visualize high dimensional image data by assigning each data point a location in a two or three dimensional map~\cite{van2008visualizing}. In our case, to concentrate on the specific features resulting in these variations in the WSI domains, the image embeddings are extracted from the penultimate layer of a pre-trained deep convolutional neural network trained on images of paintings by various artists. We chose to use the painters dataset specifically, to aid the network in distinguishing between the color spaces of different WSI domains, as well as to help us analyse the performance of our method by facilitating a meaningful visualization of the WSI clusters and the distances between them~\cite{van2010texton}.

Although the primary motivation behind our work was biomedical in nature, in order to demonstrate the effectiveness of the proposed method on natural images, we also evaluate it on Virtual Image Dataset for Illumination Transfer (VIDIT)~\cite{helou2020vidit}. We derive five domains from the illumination setting at five different (2500K, 3500K, 4500K, 5500K and 6500K) color temperatures. All the domains have 2400 samples and we use the train:test split of 9:1 at both 256 and 1024 input-output resolutions.

\begin{figure*}[ht]
\centering
\includegraphics[height=10.5cm,width=14.2cm]{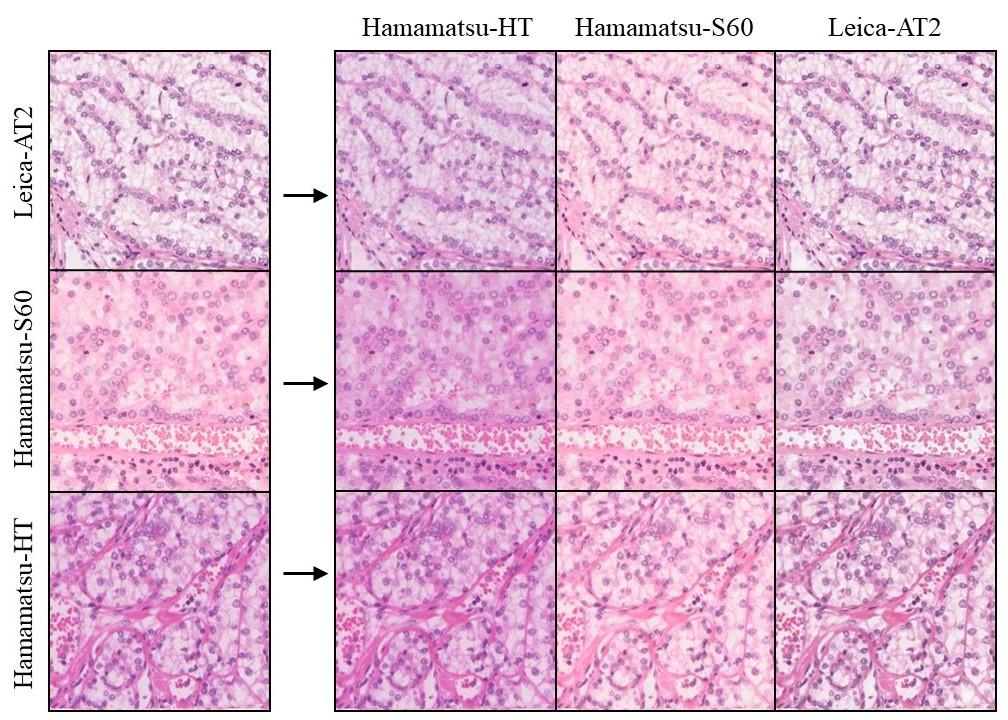}
\caption{Multi-domain translation results using MultiPathGAN on our WSI dataset for inter-domain normalization. The column on the left shows the input WSI patch while the remaining columns on the right show the translated output for the three domains as per the target label. This is achieved using a single generator.}
\Description{figure description}
\label{fig:3}
\end{figure*}

\subsection{Evaluation Metrics}

To quantify how well different methods preserve the salient structures present in the original patches, we use Peak Signal-to-Noise Ratio (PSNR), Structural Similarity (SSIM)~\cite{hore2010image}, Multi-Scale Structural Similarity Index (MS-SSIM)~\cite{wang2003multiscale} and Haar Wavelet-Based Perceptual Similarity Index (HaarPSI)~\cite{reisenhofer2018haar} measures. While PSNR depends purely on the pixel-wise differences in pixel value between the two compared images, SSIM, MS-SSIM and HaarPSI are more correlated with the perceptual differences between them. These metrics have high sensitivity to detect distortions between images as well as the differences arising due to luminance and contrast changes.

\subsection{Experimental Results}

\subsubsection{Qualitative Analysis.}

We train and evaluate MultiPathGAN on our dataset with three WSI domains. Fig.~\ref{fig:3} shows typical examples of translation results. 
As confirmed independently by a pair of experienced pathologists, observe that our method provides high quality of translation, appropriately adapting the style of images (that is the image acquisition characteristics of specific domains), while correctly preserving the salient anatomical structures (cell types and morphology, etc.).

\begin{figure}[ht]
\centering
\includegraphics[height=2.6cm,width=\linewidth]{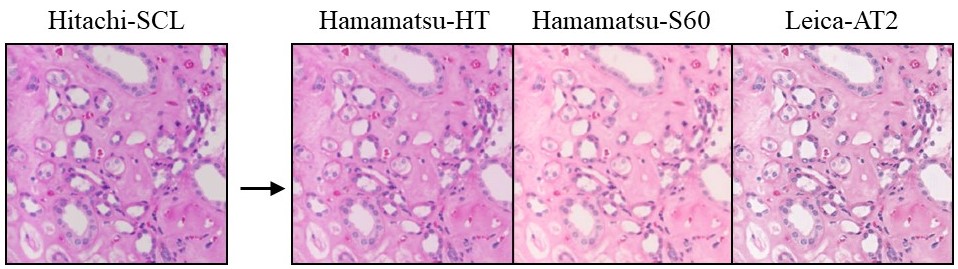}
\caption{Translations from the unseen Hitachi-SCL (leftmost column) domain to our three dataset domains used for training MultiPathGAN generator.}
\Description{figure description}
\label{fig:4}
\end{figure}

In addition, we also translate patches from the fourth unseen WSI domain to the three domains from our dataset used while training MultiPathGAN. As can be seen from Fig.~\ref{fig:4}, MultiPathGAN can effectively translate, and hence normalize, data from an alien subset of the WSI space into a known domain. This shows that normalizing unseen subsets of WSI color space using MultiPathGAN could be achieved without retraining the whole network. However, considering that this is an unseen domain, the inverse translation cannot be achieved without domain specific training, i.e.\ without the parameters which correspond to this domain, along with the corresponding target labels.

\begin{figure*}[ht]
\centering
\includegraphics[height=10.5cm,width=14.2cm]{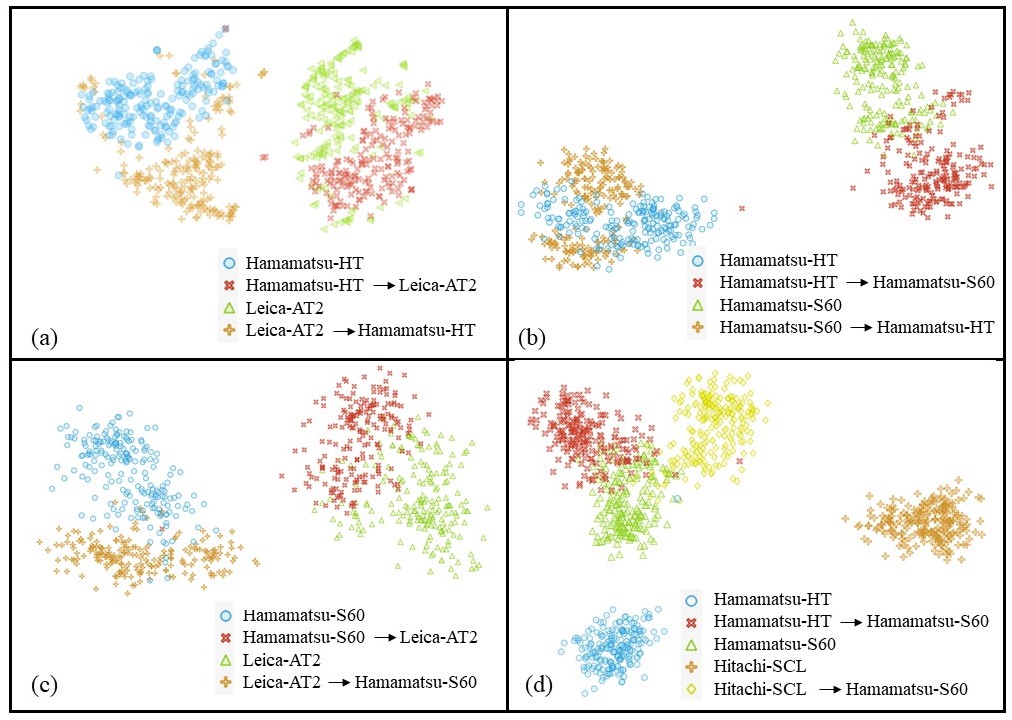}
\caption{For comprehensibility, we only show TSNE representation for real and translated samples between two domains at a time  (a)(b)(c).  We also show the effectiveness of MultiPathGAN at translating unseen domain (Hitachi-SCL) to one of the domains used during the training phase (d).}
\Description{figure description}
\label{fig:5}
\end{figure*}

We also use TSNE to visualize the effect of our translation upon the normalization of WSI patches in the deep neural network feature space. 
From Fig.~\ref{fig:5} it can be seen that the patches normalized using MultiPathGAN match well with the distinctive distribution of the target class.
This can provide an intuition for the improved performance in terms of evaluation metrics as discussed in the next subsection. In addition to Fig.~\ref{fig:4}, we can also use Fig.~\ref{fig:5}(d) to extrapolate the effectiveness of MultiPathGAN in translating an unseen WSI domain into a known WSI domain without the need to retrain the entire network.

\begin{figure*}[ht]
\centering
\includegraphics[height=3.5cm,width=17.8cm]{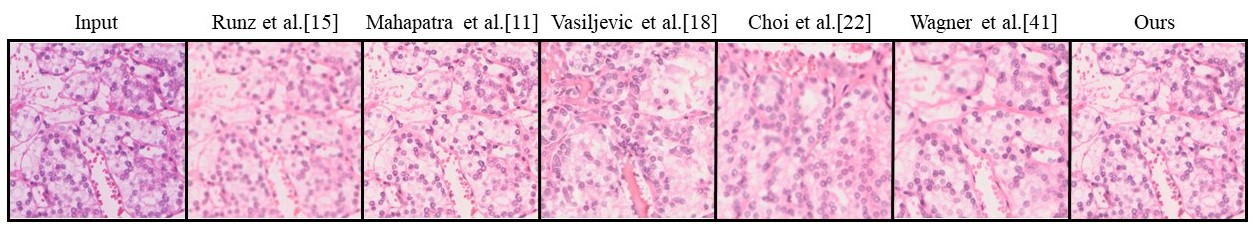}
\caption{Comparison with previous methods when translating from Hamamatsu-HT $\rightarrow$ Hamamatsu-S60.}
\Description{figure description}
\label{fig:6}
\end{figure*}

Finally, in Fig.~\ref{fig:6} we compare our cross-domain MultiPathGAN to the existing methods in the literature and observe that our method not only provides high quality and perceptually meaningful WSI normalization results, but also that it can normalize multiple WSI domains simultaneously, using a single generator in a single training session. While the method of Runz et al.~\cite{runz2021normalization} effectively translates the style between two domains, it fails at preserving the quality of the final image and resulting in blur. Mahapatra et al.~\cite{mahapatra2020structure} have shown that using semantic guidance in addition to adversarial and cycle reconstruction loss in a CycleGAN, we can preserve the detailed structural information, however, their method is restricted to only two WSI domains. In their attempt, Vasilejic et al.~\cite{vasiljevic2021towards} have used StarGAN to translate between multiple staining modalities, but when we consider stain variations in a single modality, it fails at preserving the fine structure of the tissue components in the translated output. In addition  to the StarGAN, we also show results from its newer variant by Choi et al.~\cite{choi2020stargan} and found that the generator produces images that are perceptually valid but slightly blurry and completely out of context with regards to the input WSI tissue structure. Translated samples from Wagner et al.~\cite{wagner2021structure} look realistic when compared to StarGAN but it struggles at preserving the detailed structural information and introduces objects that are out of context from the original image.

\subsubsection{Quantitative Analysis.}

For the quantitative evaluation, we first report the performance metrics for translations between two domains, namely those of the generated image $G(x, y_{trg})$ and of the original image $x$, for different methods in Table~\ref{table:1}. We compute the evaluation metrics between a real image and the same image translated to other domains in all comparative assessments. As can be seen, our model performs well with all the metrics, indicating that our model produces the most realistic WSI patches by preserving the fabrication of the tissue as well as normalizing the stains. These outputs match the style of the target domain and preserve the structure of the input patch, thereby confirming and corroborating the findings of our previous qualitative analysis shown in Fig.~\ref{fig:6}.

\setlength{\tabcolsep}{4pt}
\begin{table}
\begin{center}
\caption{Comparison with previous methods. We calculate PSNR, SSIM, MS-SSIM and HaarPSI between the test set of Hamamatsu-HT patches and Hamamatsu-HT $\rightarrow$ Hamamatsu-S60 translations.}\vspace{-5pt}
\label{table:1}
\begin{tabular}{lllll}
\hline\noalign{\smallskip}
Method & \hfil PSNR & \hfil SSIM & \hfil MS-SSIM & \hfil HaarPSI\\
\noalign{\smallskip}
\hline
\noalign{\smallskip}
Runz et al.~\cite{runz2021normalization}  & \hfil $21.16$ & \hfil $0.80$ & \hfil $0.92$ & \hfil $0.77$\\
Mahapatra et al.~\cite{mahapatra2020structure} & \hfil $21.96$ & \hfil $0.93$ & \hfil $0.96$ & \hfil $0.88$\\
Vasiljevic et al.~\cite{vasiljevic2021towards} & \hfil $14.73$ & \hfil $-0.20$ & \hfil $0.01$ & \hfil $0.53$\\
Choi et al.~\cite{choi2020stargan} & \hfil $16.19$ & \hfil $0.20$ & \hfil $0.15$ & \hfil $0.38$\\
Wagner et al.~\cite{wagner2021structure} & \hfil $18.92$ & \hfil $0.58$ & \hfil $0.64$ & \hfil $0.55$\\
MuliPathGAN (Ours) & \hfil $\textbf{22.96}$ & \hfil $\textbf{0.96}$ & \hfil $\textbf{0.98}$ & \hfil $\textbf{0.93}$\\
\hline
\end{tabular}
\end{center}
\vspace{-10pt}
\end{table}
\setlength{\tabcolsep}{1.4pt}

To evaluate our model on the unseen data, we translate between the three test domains including the additional unseen class and compare them with each other using the similarity metrics. From Table~\ref{table:2} we can notice that our model performs well with the domain transfer between the test data for all the cases. However, when we translate an unseen domain (i.e.\ HitachiSCI in this case) to one of the domains used at the time of training MultiPathGAN, there is a slight decrease in the performance in terms of the quantitative evaluation metrics. This difference is due to the lack of knowledge about the distance of the parameter distribution of our trained network from the distribution of the unseen domain and would remain constant for all such domains, provided they come from a similar locus. Nevertheless, these translations for unseen domains still deliver a good result and increasing the number of data domains to train MultiPathGAN would make the generator invariant to these unseen cohorts. This in return would further improve the performance  of MultiPathGAN at normalizing unseen WSI domains.

\setlength{\tabcolsep}{4pt}
\begin{table}
\begin{center}
\caption{Comparison of translation performance of MultiPathGAN between seen and unseen domain pairs.}\vspace{-5pt}
\label{table:2}
\begin{tabular}{lllll}
\hline\noalign{\smallskip}
Domain Transfer & PSNR & SSIM & MS-SSIM & HaarPSI\\
\noalign{\smallskip}
\hline
\noalign{\smallskip}
HT-to-S60  & \hfil $22.96$ & \hfil $0.96$ & \hfil $0.98$ & \hfil $0.94$\\
AT2-to-S60 & \hfil $27.21$ & \hfil $0.95$ & \hfil $0.95$ & \hfil $0.96$\\
S60-to-HT & \hfil $26.98$ & \hfil $0.94$ & \hfil $0.95$ & \hfil $0.95$\\
AT2-to-HT & \hfil $35.69$ & \hfil $0.97$ & \hfil $0.99$ & \hfil $0.98$\\
S60-to-AT2 & \hfil $22.63$ & \hfil $0.96$ & \hfil $0.97$ & \hfil $0.93$\\
HT-to-AT2 & \hfil $36.47$ & \hfil $0.98$ & \hfil $0.99$ & \hfil $0.98$\\
\hline
SCL-to-S60 & \hfil $18.50$ & \hfil $0.90$ & \hfil $0.94$ & \hfil $0.85$\\
SCL-to-HT& \hfil $23.77$ & \hfil $0.90$ & \hfil $0.93$ & \hfil $0.87$\\
SCL-to-AT2 & \hfil $19.32$ & \hfil $0.88$ & \hfil $0.94$ & \hfil $0.87$\\
\hline
\end{tabular}
    \begin{tablenotes}
        \small
        \item Abbreviations: HT: Hamamatsu-HT, S60: Hamamatsu-S60, \\
        AT2: Leica-AT2 and SCL: Hitachi-SCL
    \end{tablenotes}
\end{center}
\vspace{-10pt}
\end{table}
\setlength{\tabcolsep}{1.4pt}

\subsubsection{Analysis of Individual Loss Components.}

\begin{figure*}[ht]
\centering
\includegraphics[height=3.5cm,width=16.5cm]{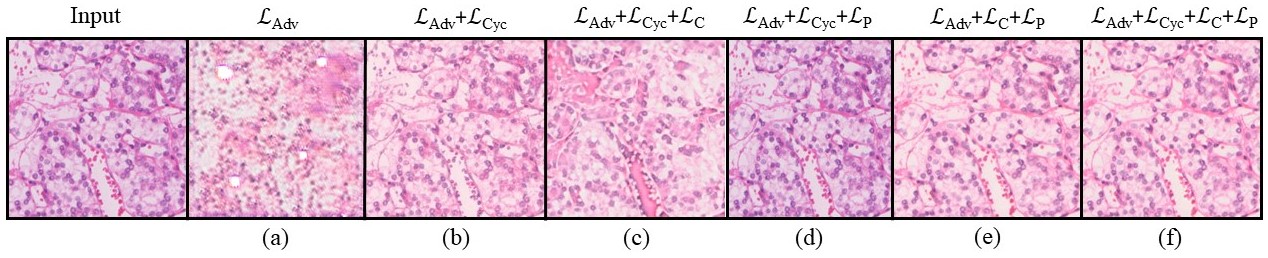}
\caption{Visual comparison of different loss functions when translating from Hamamatsu-HT $\rightarrow$ Hamamatsu-S60.}
\Description{figure description}
\label{fig:7}
\end{figure*}

To evaluate the effect of individual loss components in our network,  we add them successively to our general baseline architecture while training on our dataset. We illustrate their effects in Fig.~\ref{fig:7}, where we normalize the input domain of our test data by translating it to the other domain (Hamamatsu-HT $\rightarrow$ Hamamatsu-S60). Moreover, Table~\ref{table:3} provides the results in terms of our chosen evaluation metrics for these configurations. We start with our baseline architecture consisting of our Generator $G$ with a ResNet backbone, a PatchGAN discriminator $D$ and an adversarial loss $\mathcal{L}_{Adv}$.

\begin{figure}[ht]
\centering
\includegraphics[width=\linewidth]{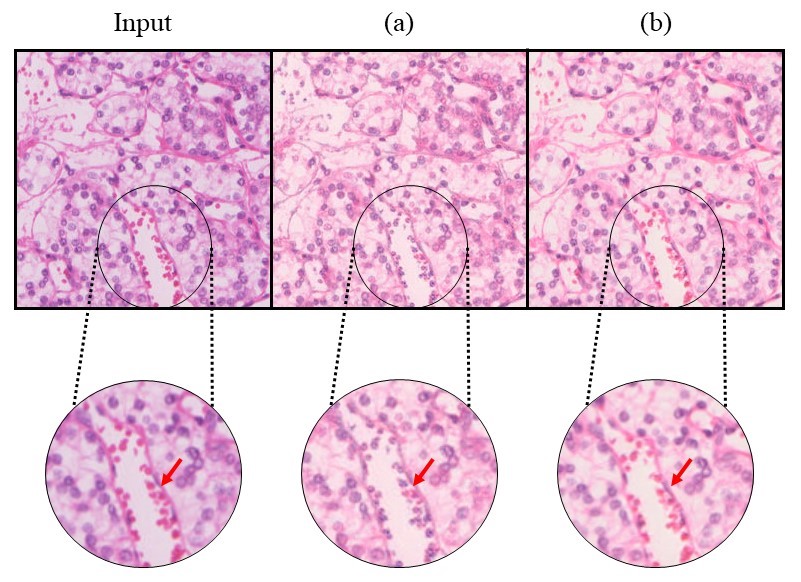}
\caption{Effect of adding Perception loss $\mathcal{L}_{P}$ (b) when compared to Reconstruction loss $\mathcal{L}_{Cyc}$ (a).}
\Description{figure description}
\label{fig:8}
\end{figure}

As can be seen from Fig.~\ref{fig:7}(a) and Table~\ref{table:3}, and as expected from previous work, using adversarial loss undeniably helps translate the style of the input to the target output.  However, it is also clear that without any semantic supervision it cannot produce a result which retains the correct semantic content, that is the relevant anatomical, structural information in the present case.
Adding semantic guidance using reconstruction loss produces images whereby the generated output image correlates well with the input data, as can be seen in Fig.~\ref{fig:7}(b) and Table~\ref{table:3}; nevertheless, it fails to preserve the fine-grained semantic information. This effect can be seen in Fig.~\ref{fig:8}, where we show the generator producing images with artifacts which do not exist in their corresponding input images when we only use reconstruction loss. Although adding classification loss forces the network to classify the input image into the corresponding target image which aids at rendering it into the style of the  target domain, it again fails at preserving the structure of the input image; see Fig.~\ref{fig:7}(c) and Table~\ref{table:3}. However, the effectiveness of classification loss can be observed in its absence in Fig.~\ref{fig:7}(d) where we can see the unified effect of reconstruction and perception loss. The resultant images highly correlate with the input in terms of both content and style, resulting in the least perceptual distance (see Table~\ref{table:3}), therefore demonstrating that the network fails to translate it to the target domain without classification loss. 

\begin{figure*}[!t]
\centering
\includegraphics[height=3.5cm,width=16.5cm]{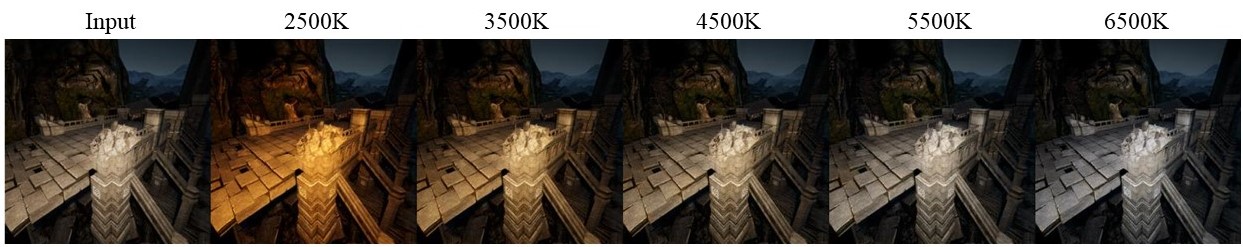}
\caption{Translation results using MultiPathGAN trained on five different color temperature domains from VIDIT.}
\Description{figure description}
\label{fig:9}
\end{figure*}

We also train MultiPathGAN in the absence of reconstruction loss and observe a good performance with the generator trained on perception loss producing images that display tissue structures identical to the input in terms of fine-grained features, as can be seen in Fig.~\ref{fig:7}(d). However, as we can observe from Table~\ref{table:3}, adding reconstruction loss preserves global semantic composition of the output image unnoticeable to the naked eye, resulting in better performance in terms of the evaluation metrics. Lastly, we observe that using the final MultiPathGAN loss produces images which are semantically indistinguishable from the input image (see Table~\ref{table:3}) while at the same time also closer to the target domain in terms of its style composition, as illustrated by the example in Fig.~\ref{fig:7}(f). 

\setlength{\tabcolsep}{4pt}
\begin{table}
\begin{center}
\caption{Comparison of different MultiPathGAN losses.}\vspace{-5pt}
\label{table:3}
\begin{tabular}{lllll}
\hline\noalign{\smallskip}
Loss & \hfil PSNR & \hfil SSIM & \hfil MS-SSIM & \hfil HaarPSI\\
\noalign{\smallskip}
\hline
\noalign{\smallskip}
$\mathcal{L}_{Adv}$  & \hfil $18.20$ & \hfil $0.39$ & \hfil $0.49$ & \hfil $0.45$\\
$\mathcal{L}_{Adv}+\mathcal{L}_{Cyc}$ & \hfil $22.04$ & \hfil $0.85$ & \hfil $0.93$ & \hfil $0.90$\\
$\mathcal{L}_{Adv}+\mathcal{L}_{Cyc}+\mathcal{L}_{C}$ & \hfil $14.05$ & \hfil $-0.22$ & \hfil $0.03$ & \hfil $0.56$\\
$\mathcal{L}_{Adv}+\mathcal{L}_{Cyc}+\mathcal{L}_{P}$ & \hfil $37.76$ & \hfil $0.97$ & \hfil $0.99$ & \hfil $0.99$\\
$\mathcal{L}_{Adv}+\mathcal{L}_{C}+\mathcal{L}_{P}$ & \hfil $22.79$ & \hfil $0.93$ & \hfil $0.97$ & \hfil $0.91$\\
$\mathcal{L}_{Adv}+\mathcal{L}_{Cyc}+\mathcal{L}_{C}+\mathcal{L}_{P}$ & \hfil $22.96$ & \hfil $0.96$ & \hfil $0.98$ & \hfil $0.93$\\
\hline
\end{tabular}
\end{center}
\vspace{-10pt}
\end{table}
\setlength{\tabcolsep}{1.4pt}

\subsubsection{Qualitative results on VIDIT}
Finally, we demonstrate empirically that our method can maintain structural consistency between the original images and transferred images irrespective of the discrepancy in domain composition and style. As can be seen from Fig.~\ref{fig:9}, MultiPathGAN can synthetically render a scene as if re-illuminated using an unseen lighting setup,
which is useful in a variety of tasks, for example in reference-based image relighting. 
More results demonstrating the effectiveness of MultiPathGAN and its robustness to lighting conditions on VIDIT can be found in the supplementary material.


\section{Discussion}

Considering the widespread use of convolutional neural networks in CAD systems, stain normalization  plays a crucial part in the interoperability and improved accuracy of their recommendations and results. MultiPathGAN provides a reliable approach to remove variations in the acquired images by providing clinically meaningful output images, normalized at the pixel resolution comparable to real biomedical images, for further assessment. There are several reasons for the improved robustness, flexibility, and accuracy of MultiPathGAN in WSI normalization when compared to all previously proposed methods. First, unlike these, MultiPathGAN learns the mapping from the entire dataset instead of relying on a single reference image, thus avoiding the complexity of choosing reference images. Second, most existing methods that use adversarial learning methods focus on using GANs to translate only between two domains, thus limiting the potential of GANs to translate between multiple domains simultaneously using a single generator and style code~\cite{choi2018stargan}. Moreover, all existing methods are ineffective, in that even if multiple generators are used to translate between different domain pairs, the learning of the global features available to each generator is done separately. Hence, the entirety of information available within the training data corpus is not used to its full extent. This inability to fully utilize the training data by jointly training the domains inevitably limits the quality of generated samples and the ability of the generator to successfully normalize unseen WSI domains. Therefore, in MultiPathGAN we can say that the shared data from each domain helps to learn domain-invariant features which produces a regularization effect, thus facilitating better generalization to unseen WSI samples.

Lastly, we analysed different loss functions and introduced a pre-trained auxiliary ResNet feature extraction network to calculate a perceptual distance. This distance is employed in the additional perceptual loss function used while training the generator to preserve fine details of the input WSI patches in the normalized output. The smaller features are better propagated to higher convolutional levels in residual networks owing to skip connections, such that, there is an implicit weighing between fine and large features. We use the activations from the final layers of the ResNet feature extraction network to enable us to retain the finer details of the original image during the training process. This helps us to produce images which preserve the fine structure of the input image while effectively translating to the style of the target image domain.

The applications of MultiPathGAN are not limited to WSI stain normalization. Indeed, any application wherein there is high value in the exact semantic content of images can benefit from the method's ability to preserve such content while translating between multiple, even unseen domains.

\section{Conclusions}
To summarise, we presented the state-of-the-art adversarial model, MultiPathGAN, an unsupervised multi-domain image-to-image translation network that outperforms all other stain normalization methods devised for histopathological whole slide imagery. Our main contributions are: (1) domain invariance, whereby we can easily expand the set of domains (including to unseen ones) for normalization using only a single generator-discriminator, and (2) compelling the output image to preserve the salient anatomical tissue structure of the input image using an auxiliary feature extraction network and perception loss. We also demonstrated that our model can be readily adopted in other image-to-image translation applications in which it is important to preserve the structural content of the input image while only changing its style. In closing, MultiPathGAN has the potential to enhance the performance of medical image based diagnostic systems wherein annotations are not readily available, and its key strength stems from a meaningful constrainment of adversarial networks and their tendency to induce unwanted diversity which is manifest in undesirable image artifacts.

\section{Supplementary Material}
\subsection*{Additional Qualitative Results}

\begin{figure*}[ht]
\centering
\includegraphics[height=19.6cm,width=13.5cm]{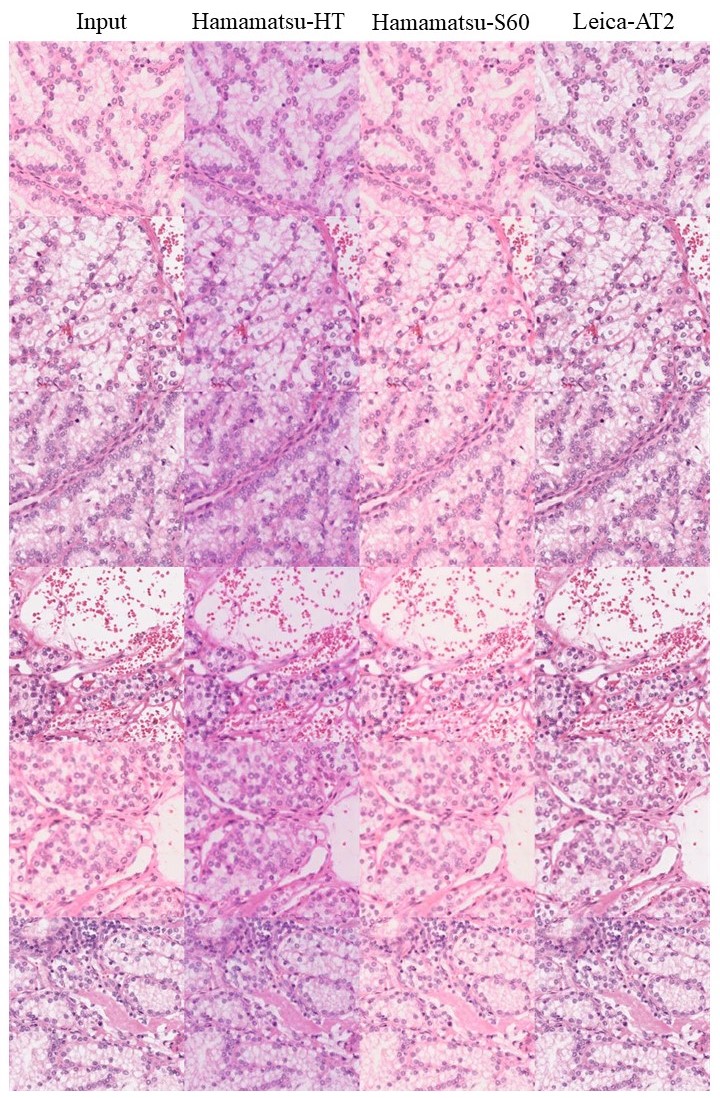}
\caption{Additional translation results using MultiPathGAN on our WSI dataset for inter-domain normalization. The column on the left shows the random input WSI patch while the remaining columns on the right show the translated output for the three domains as per the target label.}
\label{additional_wsi_eg}
\end{figure*}

\begin{figure*}[ht]
\centering
\includegraphics[height=19.5cm,width=15.9cm]{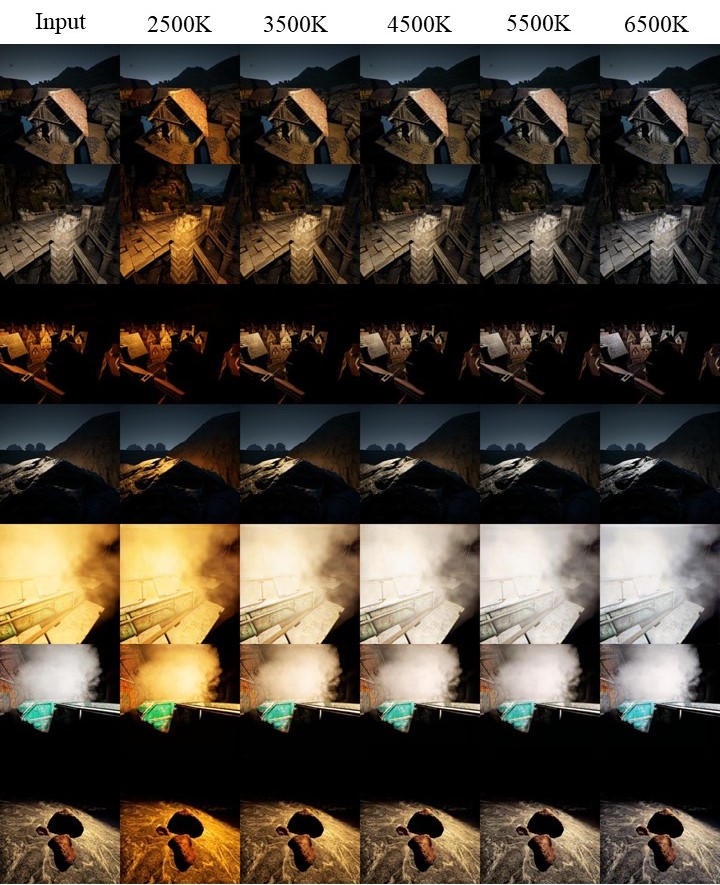}
\caption{Additional translation results using MultiPathGAN trained on five different color temperature domains from VIDIT at $256\times256$ resolution. The column on the left shows the input scene at random color temperature while the remaining columns on the right show the translated output for different illumination temperatures.}
\label{additional_vidit_eg}
\end{figure*}

\begin{figure*}[ht!]
\centering
\includegraphics[height=12.5cm,width=15.9cm]{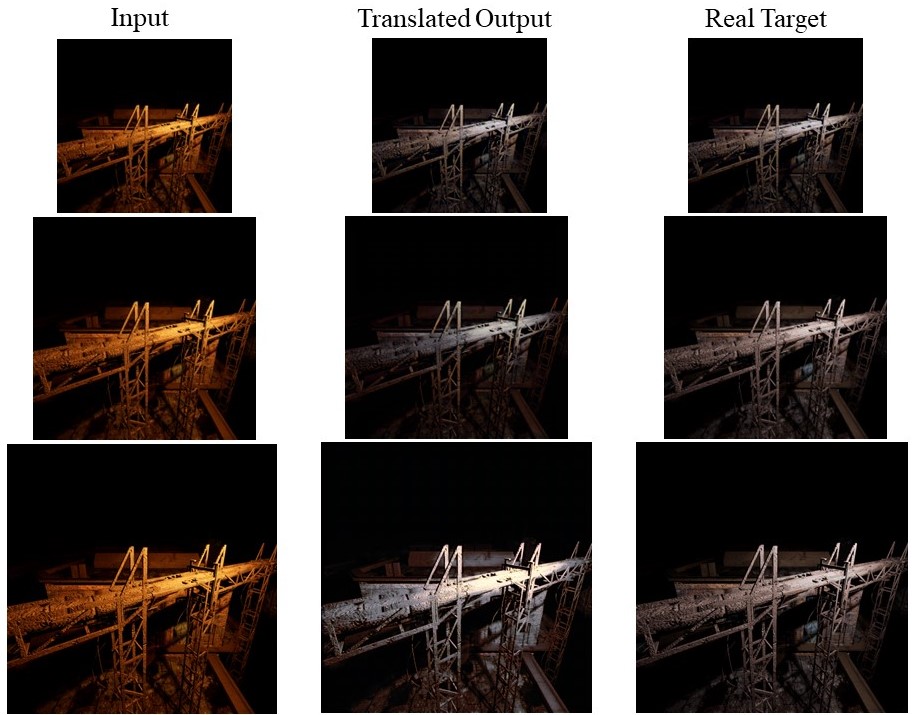}
\caption{Translation results between 2500K and 6500K color temperatures for VIDIT dataset generated at $256\times256$, $512\times512$ and $1024\times1024$ resolutions.}
\label{Vidit_256_1024}
\end{figure*}

 Fig.~\ref{additional_wsi_eg} and Fig.~\ref{additional_vidit_eg} show additional images generated using MultiPathGAN at $256\times256$ resolution for stain normalization on our kidney tissue WSI dataset and scene re-illumination on VIDIT dataset, respectively. Fig.~\ref{Vidit_256_1024}, shows translation results for VIDIT dataset generated at $256\times256$, $512\times512$ and $1024\times1024$ resolutions.

\subsection*{Additional Quantitative Results}

We compare the performance metrics for MultiPathGAN trained on images at $256\times256$, $512\times512$ and $1024\times1024$ resolutions from the VIDIT dataset. We compute these metrics between the translated output and the real image at the target color temperature as depicted in Fig.~\ref{Vidit_256_1024}. As can be seen from Table ~\ref{Metric_eval_dims}, our model performs better according to all metrics at $256\times256$ resolution. We also compare the performance of MultiPathGAN trained at different $\lambda_{P}$ values on our WSI dataset in Table ~\ref{Metric_eval_lambdaP}.

\begin{table}[ht!]
\begin{center}
\caption{Comparison of translation performance of MultiPathGAN trained at $256\times256$, $512\times512$ and $1024\times1024$ resolutions using VIDIT dataset.}
\label{Metric_eval_dims}
\begin{tabular}{lllll}
\hline\noalign{\smallskip}
I/O Dimension & \hfil PSNR & \hfil SSIM & \hfil MS-SSIM & \hfil HaarPSI\\
\noalign{\smallskip}
\hline
\noalign{\smallskip}
\hfil $256 \times 256$ & \hfil $\textbf{30.67}$ & \hfil $\textbf{0.94}$ & \hfil $\textbf{0.98}$ & \hfil $\textbf{0.88}$\\
\hfil $512 \times 512$ & \hfil $24.67$ & \hfil $0.86$ & \hfil $0.94$ & \hfil $0.78$\\
\hfil $1024 \times 1024$ & \hfil $20.90$ & \hfil $0.73$ & \hfil $0.92$ & \hfil $0.63$\\
\hline
\end{tabular}
\end{center}
\end{table}

\begin{table}[ht!]
\begin{center}
\caption{Comparison of translation performance of MultiPathGAN trained at various values of $\lambda_{P}$.}
\label{Metric_eval_lambdaP}
\begin{tabular}{lllll}
\hline\noalign{\smallskip}
\hfil$\lambda_{P}$ & \hfil PSNR & \hfil SSIM & \hfil MS-SSIM & \hfil HaarPSI\\
\noalign{\smallskip}
\hline
\noalign{\smallskip}
$0.25$ & \hfil $22.60$ & \hfil $0.95$ & \hfil $0.98$ & \hfil $0.92$\\
$0.50$ & \hfil $22.54$ & \hfil $0.95$ & \hfil $0.98$ & \hfil $0.92$\\
$0.75$ & \hfil $\textbf{22.96}$ & \hfil $\textbf{0.96}$ & \hfil $\textbf{0.98}$ & \hfil $\textbf{0.93}$\\
$1.00$ & \hfil $22.67$ & \hfil $0.95$ & \hfil $0.98$ & \hfil $0.93$\\
\hline
\end{tabular}
\end{center}
\end{table}

\subsection*{Mean Color Difference}
To measure the color difference between the input image, output image and the target image, we calculate the difference of euclidean distances between their corresponding pixels in the CIELAB color space.~\cite{connolly1997study, hill1997comparative}. The resultant mean difference with higher values indicates a better match in terms of color value between the generated fake image and the target image. This value for MultiPathGAN comes out to be -0.0036 followed closely by Vasiljevic et al.~\cite{vasiljevic2021towards}[-0.0039]. For Runz et al.~\cite{runz2021normalization}, Mahapatra et al.~\cite{mahapatra2020structure} and Wagner et al.~\cite{wagner2021structure} this metric value was -0.0054, -.0055 and -0.0056 respectively.

\balance


\bibliographystyle{ACM-Reference-Format}
\bibliography{main.bib}

\end{document}